\documentclass[12pt]{article}
\usepackage[utf8]{inputenc}
\usepackage[english]{babel}
\usepackage[margin=1in]{geometry}
\usepackage{amsmath, amssymb}
\usepackage{csquotes}            
\usepackage[
  backend=biber,                
  style=nature,                 
  sorting=none,                 
  maxbibnames=6,                
  minbibnames=1                 
]{biblatex}
\usepackage{setspace}
\usepackage{titlesec}
\usepackage{graphicx}
\usepackage{caption}
\usepackage{enumitem}
\usepackage{subcaption} 
\usepackage{float}
\usepackage{booktabs}
\usepackage{hyperref}
\usepackage{comment}
\usepackage{xcolor}
\title{Epidemiological dynamics with clinically-derived infectiousness and incubation time courses}
\author{
   Miguel A. Cajahuanca Ricaldi  \and Yaroslav Ispolatov \\
    \textit{Department of Physics, University of Santiago, Chile} \\
    \texttt{miguel.cajahuanca@usach.cl}, \texttt{jaros007@gmail.com}
}
\date{\today}

\titleformat{\section}{\normalfont\large\bfseries}{\thesection.}{1em}{}
\titleformat{\subsection}{\normalfont\normalsize\bfseries}{\thesubsection.}{1em}{}

\begin{document}
\maketitle
\begin{abstract}
To better predict the dynamics of epidemics such as COVID-19, it is important not only to investigate the network of local and long-range contagious contacts but also to understand the temporal dynamics of infectiousness and detectable symptoms. Here, we present a model of infection spread in a well-mixed group of individuals, which usually corresponds to a node in large-scale epidemiological networks. The model uses delay equations that take into account the duration of infection and are based on experimentally derived time courses of viral load and shedding, as well as the detectability of symptoms. We show that due to an early onset of infectiousness, which is reported to be synchronous or even precede the onset of detectable symptoms, the tracing and immediate testing of all who came in contact with the detected infected individual reduce the spread of epidemics, hospital load, and fatality rate. We also investigate how the strictness and promptness of the isolation of infected individuals affect the outcome of epidemics. We hope that these more precise node dynamics could be incorporated into complex large-scale epidemiological models to improve the accuracy and credibility of predictions. 
\end{abstract} 

\section{Introduction}

The COVID-19 pandemic demonstrated that, although globalization of this century has favored an  unprecedented in human history interconnectedness, it has also significantly increased our vulnerability to the spread of infectious diseases. In this context, the effective implementation of public health measures plays a fundamental role, while a lack of ability to mitigate the advance of pandemics entails serious health consequences, as well as economic repercussions.

For public health measures to be effective, it is crucial to understand the specific dynamics of the virus. One of the distinguishing features of COVID-19 is that an infected patient may be infectious before showing symptoms; see, e.g., \cite{he2020temporal}. In addition, the proportion of asymptomatic patients has been shown to be considerable, contributing significantly to the silent transmission of the virus. These characteristics make early detection of infection difficult and result in silent circulation of the virus, significantly complicating the design of effective interventions. For these peculiar characteristics of COVID-19, classical models such as SIR or SEIR, which assume that transmission of contagion begins once the individual enters the infectious phase - implying that the ability to infect is associated with the onset of symptoms - do not prove to be adequate, as shown in \cite{10.1093/aje/kwt133}, \cite{doi:10.1137/S0036144500371907} and \cite{LLOYD200159}.

In this work, we develop a more accurate epidemic model that incorporates presymptomatic transmission dynamics, addressing the limitations of classical compartmental models. Building on the integro-differential framework introduced in \cite{kermack1927contribution}, we propose a modified model incorporating infectivity and incubation period profiles derived from empirical clinical data. Our model is based on three time-dependent characteristics of infection, each treated as a normalized probability density function estimated directly from clinical data. The first, \textit{the intrinsic generation time distribution}, $G_{I}(t)$, represents the probability density that a primary case transmits infection to a secondary individual exactly $t$ days after the primary case’s own infection. It is generally different from the \textit{realized household generation time distribution}, $G_{H}(t)$, the probability density that a primary case infects another member of the same household exactly $t$ days after the primary case’s own infection. It captures not only the pathogen’s infectiousness profile but also the household’s contact structure. While the intrinsic generation time reflects the natural infectiousness profile of a disease in an idealized setting, the realized household generation time captures how transmission unfolds in real-world constrained environments. The third is \textit{the incubation period distribution}, $P(t)$, the probability density that an infected individual develops symptoms exactly $t$ units of time after infection. 

\begin{figure}[H]
    \centering
    \includegraphics[width=0.8\textwidth]{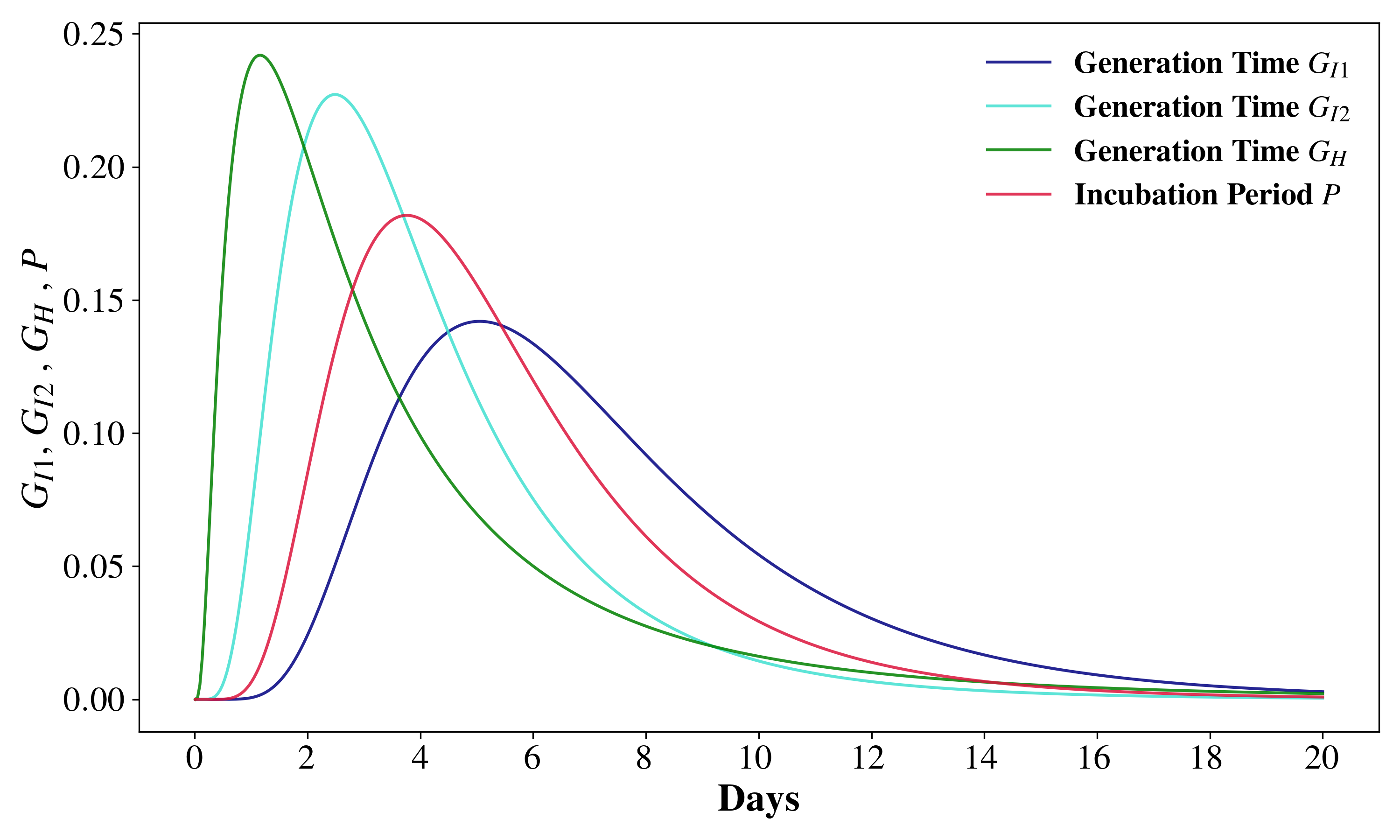}
    \caption{Distributions of  the generation times $G_{I1}, G_{I2}, G_{H}$ and of the incubation period $P$.}
    \label{fig:figure1}
\end{figure}

We consider threeg distributions of generation times, each estimated by statistical inference from clinical data: the generation time at the onset of the COVID-19 outbreak, denoted \( G_{I1} \), corresponding to the second week of January 2020; the generation time at the end of the same month, denoted \( G_{I2} \) \cite{chen2022inferring}; and the household generation time, \( G_{H} \), reported in \cite{10.7554/eLife.70767}. For the incubation period distribution, we follow the results of \cite{ejima2022incubation}, where these times were inferred from viral load data. Figure~\ref{fig:figure1} displays the probability distributions of these four characteristic times.

\section{Methods}

\subsection{Population dynamics}

We consider the dynamics of a well-mixed subpopulation partitioned into Susceptible~(S), Infected~(I), Quarantined~(Q), Recovered~(R), and Deceased~(D) compartments. Let \(S(t)\) denote the fraction of individuals that are susceptible at time \(t\). Susceptible individuals become infected through contact with infectious people, whether they are in the Infected or Quarantined compartment, regardless of whether their infection has already been detected.

The quarantine compartment, \( Q(t) \), represents the fraction of the population that, after having shown symptoms that were detected, is being isolated either at home or in hospitals.
\begin{figure}[h]
    \centering
    \includegraphics[width=1.0\linewidth]{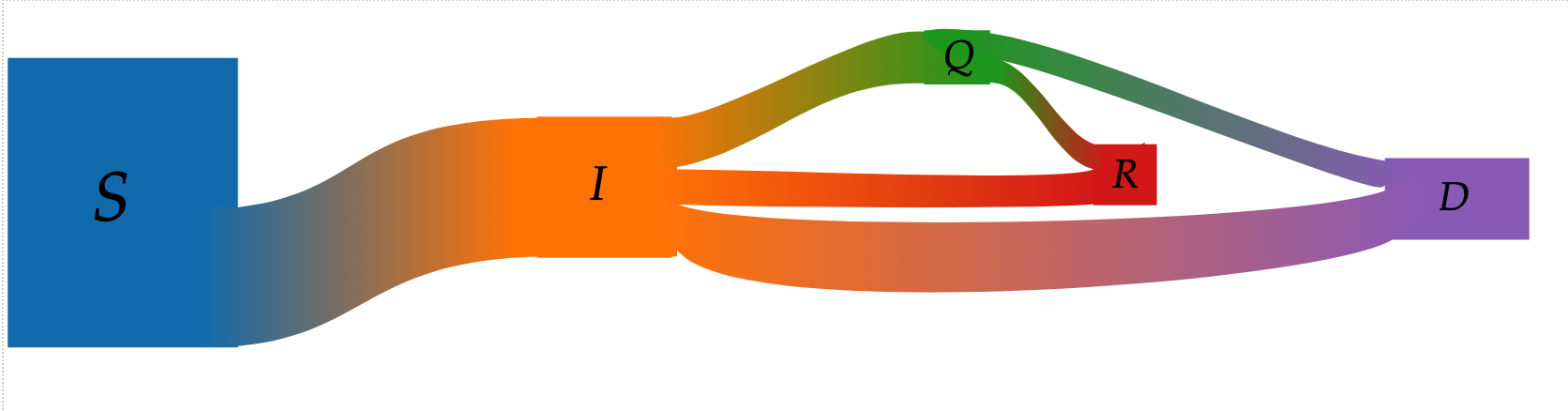}
    \caption{flow diagram of the model}
    \label{fig:enter-label}
\end{figure}

The concentrations of undetected and quarantined infected individuals are denoted by \( I(\tau, t) \) and \( Q(\tau, t) \), respectively, where \( \tau \), with \( 0 \leq \tau \leq t \), represents the time at which the individual became infected. 
To model the reduced infectiousness of quarantined individuals, we define the parameter $\theta \in [0,1]$ as the \textit{residual transmissibility under quarantine}, due to the reduction in contacts resulting from the implementation of isolation protocols. It represents the relative capacity of quarantined individuals to generate new infections compared to undetected, non-isolated cases. A value of $\theta = 0$ implies fully effective quarantine (no secondary infections), whereas $\theta = 1$ corresponds to no reduction in transmission. Intermediate values reflect partial effectiveness.

Quantitatively, the infection of susceptibles occurs at the rate
\begin{equation}
\frac{dS(t)}{dt} = -C_IS(t) \int_0^t\left[ G_I(t-\tau) I(\tau, t)+ \theta G_{H}(t-\tau) Q(\tau, t)\right]d\tau,
\label{eq1}
\end{equation}
where $C_I$ is the infectiousness coefficient. 

The distribution functions of the intrinsic generation time, \( G_I(t-\tau) \), and the household generation time, \( G_H(t-\tau) \), which depend on the duration of a particular infection \( t - \tau \), are central components of our model. They characterize how contagious an individual is at a given stage of infection for undetected and quarantined cases, respectively. For diseases transmitted via the nasal route, estimating the generation time distributions \( G_I(t-\tau) \) and \( G_H(t-\tau) \) poses a major challenge, since the exact timing of primary and secondary infections is often unknown. From a statistical standpoint, these distributions can be inferred from patients’ symptom onset dates and exposure intervals, as demonstrated in \cite{manica2023estimation} and \cite{xu2023incubation}.

The concentration of infected individuals follows Eq. (\ref{eq2}),
\begin{align}
\frac{\partial I(\tau, t)}{\partial t} &= \delta(t - \tau)\, C_I\, S(t) \int_0^t \left( G_I(t-\tau') I(\tau', t)+ \theta G_{H}(t-\tau') Q(\tau', t)\right) d\tau' \nonumber \\
&\quad - I(\tau, t) \left[ C_S\, P(t - \tau) + \delta(t - \tau - \tau^*) \right]
\label{eq2}
\end{align}
The gain in the number of undetected infected individuals, described by the first term in (2), comes from the new infections described above. The delta function $\delta(t - \tau)$ indicates that the beginning of a new infection occurs at the current time $t$. The loss term describes the detection of infection and thus the conversion of detected individuals into quarantined ones. It is assumed to happen at the rate dependent on the time elapsed from the time of infection and is given by the product of distribution of the incubation period $P(t - \tau)$ and the detection constant $C_S$.
The term $-I(\tau,t)\delta(t-\tau-\tau*)$ in Eq. (2) describes the resolution of a disease after a typical duration of $\tau^* \sim 20$ days, resulting in recovery or death of a patient.

The concentration of quarantined individuals increases via the detection of so far undetected infected individuals and decreases on the $\tau^*$th day of the disease as individuals recover or die,

\begin{equation}
\frac{\partial Q(\tau, t)}{\partial t} = I(\tau, t) C_S P(t - \tau) - Q(\tau, t) \delta(t - \tau - \tau^*).
\label{eq3}
\end{equation}

The approximation of a ``sudden recovery'' or a possible ``sudden death'' on the $\tau^*$th day of the disease does not affect the dynamics of infection as it has been observed that individuals destined either to recover or die lose infectiousness in about a week after the onset of symptoms, which is well in advance of $\tau^*$ . The approximation does slightly affect the death and recovery running statistics but not the final number of recovered and dead by the end of epidemics. We assume that the quarantined individuals receive better care and have a lower death rate than the undetected ones. This is reflected by the coefficient $\nu < 1$.

\begin{equation}
\frac{dD(t)}{dt} = C_D \delta(t - \tau - \tau^*) \left[ I(\tau, t) + \nu Q(\tau, t) \right].
\label{eq4}
\end{equation}

Individuals who do not die, recover,

\begin{equation}
\frac{dR(t)}{dt} = \delta(t - \tau - \tau^*) \left\{ I(\tau, t)[1 - C_D] + Q(\tau, t)[1 - C_D \nu] \right\}.
\label{eq5}
\end{equation}
The constants $C_I$ in (1,2), $C_S$ in (2,3), and $C_D$ in (4,5) are fitted to reproduce the empirical infection, the detection of infection (quarantining) and the death rates. 
To make the simulation equally adapted to both the analytical and tabulated form of $G_I$ , $G_H$ and $P$, the integrals in \eqref{eq1} and \eqref{eq2} are evaluated using the trapezoid rule with the discretization step of one day, which is the most common time unit in epidemiological and clinical data.

\subsection{Delayed quarantine}

As observed during the course of the pandemic, protocols for detection of infection and isolation of symptomatic individuals varied between the countries. Therefore, it is important to account for this variability to more accurately model the progress of the pandemic. To account for various isolation protocols and levels of social responsibility, we quantify this characteristic by the delay time, $t_{delay}$. Specifically, it takes $t_{delay}$ for an individual to enter quarantine after developing symptoms. A smaller $t_{delay}$ corresponds to a more strict isolation protocol and a higher level of society's responsibility. To model this situation, a slight modification of the model (\ref{eq1}-\ref{eq5}) is needed:
The term in (\ref{eq2},\ref{eq3}) that describe the isolation rate of patients who develop symptoms at time $\tau$,
\( I(\tau, t) \, C_S \, P(t  - \tau) \),
 have to be adjusted to account for $t_{delay}$, 
\( I(\tau, t - t_{delay}) \, C_S \, P(t - t_{delay} - \tau) \). It means that the individuals that become isolated at time $t$ have developed detectable symptoms at time $t-t_{delay}$.

\subsection{Simulation Framework}

The epidemic dynamics were simulated using a discrete-time, deterministic compartmental model implemented in Python. The model describes the temporal evolution of susceptible ($S$), infected ($I$), quarantined ($Q$), recovered ($R$), and deceased ($D$) fractions of a normalized population. It is assumed that the initial infected population is $I(\tau=0,t=0)=I_0\ll 1$ and the total initial population is $S(t=0)=1-I_0$, and the infected and quarantined compartments are structured by the infection age over a finite time $T_{res}$ of resolution of the disease, $0 \leq t-\tau\leq T_{res}$. In concordance with the majority of medical records, the unit of discretization was chosen to be equal to one day. 

The transmission of infection occurs via two parallel processes: undetected individuals ($I$) transmit the infection with a rate coefficient $C_I$, while quarantined individuals ($Q$) do so with a reduced infectiousness rate $\theta C_I$. Thus, the density of newly infected individuals at each time step is computed as a convolution of the infectivity kernels $G_I$ and $G_H$ with the corresponding compartments.
\begin{align}
I(\tau=t+1, t+1) &=  C_I\, S(t) \sum_{t'=0}^{T_{res}}\left[ G_I(t') I(\tau=t-t', t)+ \theta G_{H}(t') Q(\tau=t-t', t)\right];\\
\nonumber
I(\tau, t+1)& =I(\tau,t)\left[1-C_S\, P(t - \tau)\right],\:\: t-T_{res}\leq \tau \leq t.
\label{eq_discr}
\end{align}
The second equation reflects the detection of infection in a fraction of infected individuals with symptoms. In the same way, Eq. (\ref{eq3}) is converted into a finite difference equations for $Q(\tau,t)$. 
The kernels $G_I(t)$ and $G_{H}(t)$, along with the incubation period distribution $P(t)$, are inferred from the clinical data and illustrated in Fig. \ref{fig:enter-label}.
Eventually the infection gets resolved in $T_{res}$ days after its beginning. Infected and quarantined  individuals exit the infectious pools, dying or recovering:
\begin{align}
D(t+1) &= D(t)+C_D\,[I(\tau=t-T_{res},t)+ \nu Q(\tau=t-T_N,t)],\\
\nonumber
R(t+1) &= R(t)+(1-C_D)\,I(\tau=t-T_{res},t)+(1-C_D\nu) \,Q(\tau=t-T_N,t)].
\end{align}
The set of parameters used in simulations was  calibrated using the  epidemiological data \cite{chen2022inferring}, \cite{10.7554/eLife.70767}, \cite{ejima2022incubation}  and is summarized in Table~\ref{tab:epidemic-simulation-params}. 

\begin{table}[htbp]
  \centering
  \caption{Initial conditions and parameters used in the epidemic simulation}
  \label{tab:epidemic-simulation-params}
  \begin{tabular}{@{} l l c l @{}} 
    \toprule
    \textbf{Symbol} & \textbf{Description}                            & \textbf{Value} & \textbf{Units}        \\
    \midrule
    $T_{res}$           & Infectious period length (discrete steps)      & 20             & days                  \\
    $T_{fin}$         & Total simulation time                          & 200            & days                  \\
    $C_I$          & Transmission rate (undetected)                 & 2.75           & day$^{-1}$            \\
    $C_S$          & Detection rate                                 & 0.50           & day$^{-1}$            \\
    $C_D$          & Death rate                                   & 0.07           & day$^{-1}$            \\
    $\nu$           & Attenuation of death rate of  susceptibles      & 0.5           & dimensionless        \\
    $\theta$           & Relative contribution of quarantined to spread & $ [0,1]$           & dimensionless         \\
    $I_0$         & Initial infected fraction                      & $10^{-6}$      & dimensionless  \\
    $S_0$         & Initial susceptible fraction                   & $1 - 10^{-6}$  & dimensionless  \\
    $Q_0$         & Initial quarantined fraction                   & 0.0            & dimensionless  \\
    $R_0$, $D_0$ & Initial recovered / dead fraction              & 0.0            & dimensionless  \\
    $G_{I1}$          & Generation time distribution                   & array     &  day$^{-1}$  
    \\
    $G_{I2}$          & Alternative generation distribution            & array      & day$^{-1}$   \\
    $G_{H}$           & Quarantined generation distribution            & array     & day$^{-1}$    \\
    $P$            & Incubation period distribution                 & array      & day$^{-1}$    \\
    $\Delta t$     & Time step                                 & 1.0            & day                   \\
    \bottomrule
  \end{tabular}
  \captionsetup{font=small,labelfont=bf}
\end{table}

The simulations continue until a final time of $T_{fin} = 200$ days, at which point the values of the epidemiological compartments stabilize and the numbers of quarantined and undetected infected individuals approach zero. The time series of densities of all compartments is recorded for subsequent analysis.

\section{Results}
\begin{figure}[H]
    \centering
    \includegraphics[width=1.0\textwidth]{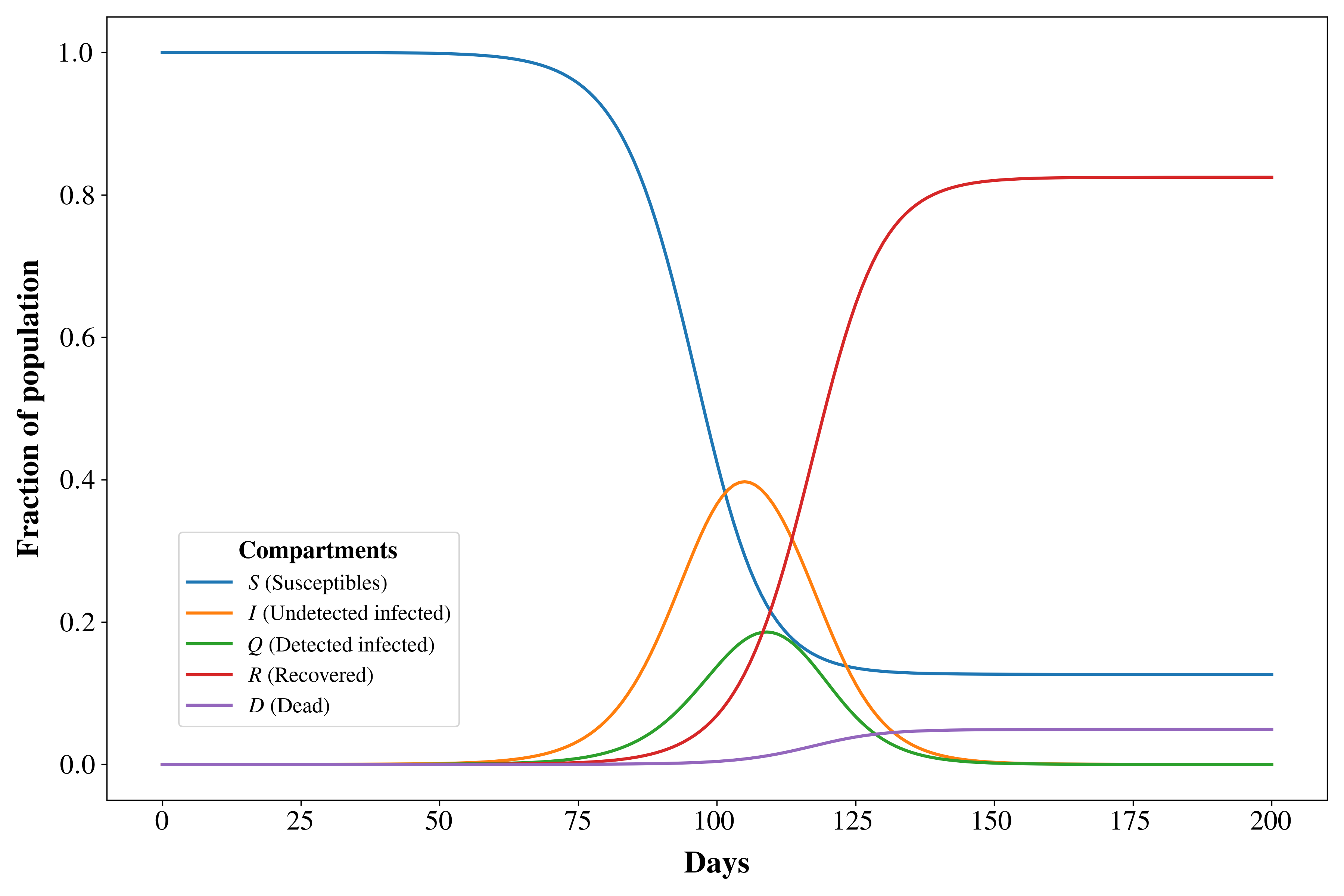}
    \caption{Temporal dynamics of fractions of dead (purple), recovered (red), undetected infected (orange), detected infected (green), and susceptible (blue) individuals for the following parameters: The infection rate coefficient $C_I = 2.75$, the coefficient in the rate of detection of infected $C_S = 0.5$, the death rate coefficient $C_D = 0.07$, the attenuation of death rate and infectiousness after the detection of the disease $\nu = 0.5$ }
    \label{fig:figure3}
\end{figure}

Figure~\ref{fig:figure3} shows epidemic time course under our model assumptions. The parameters \( C_I \), \( C_S \), and \( C_D \) were chosen to reproduce empirical features of early-stage COVID-19, specifically a doubling time of approximately 5 days and a cumulative fatality rate of around 5\%.

\begin{figure}[H]
    \centering
    \includegraphics[width=1.0\textwidth]{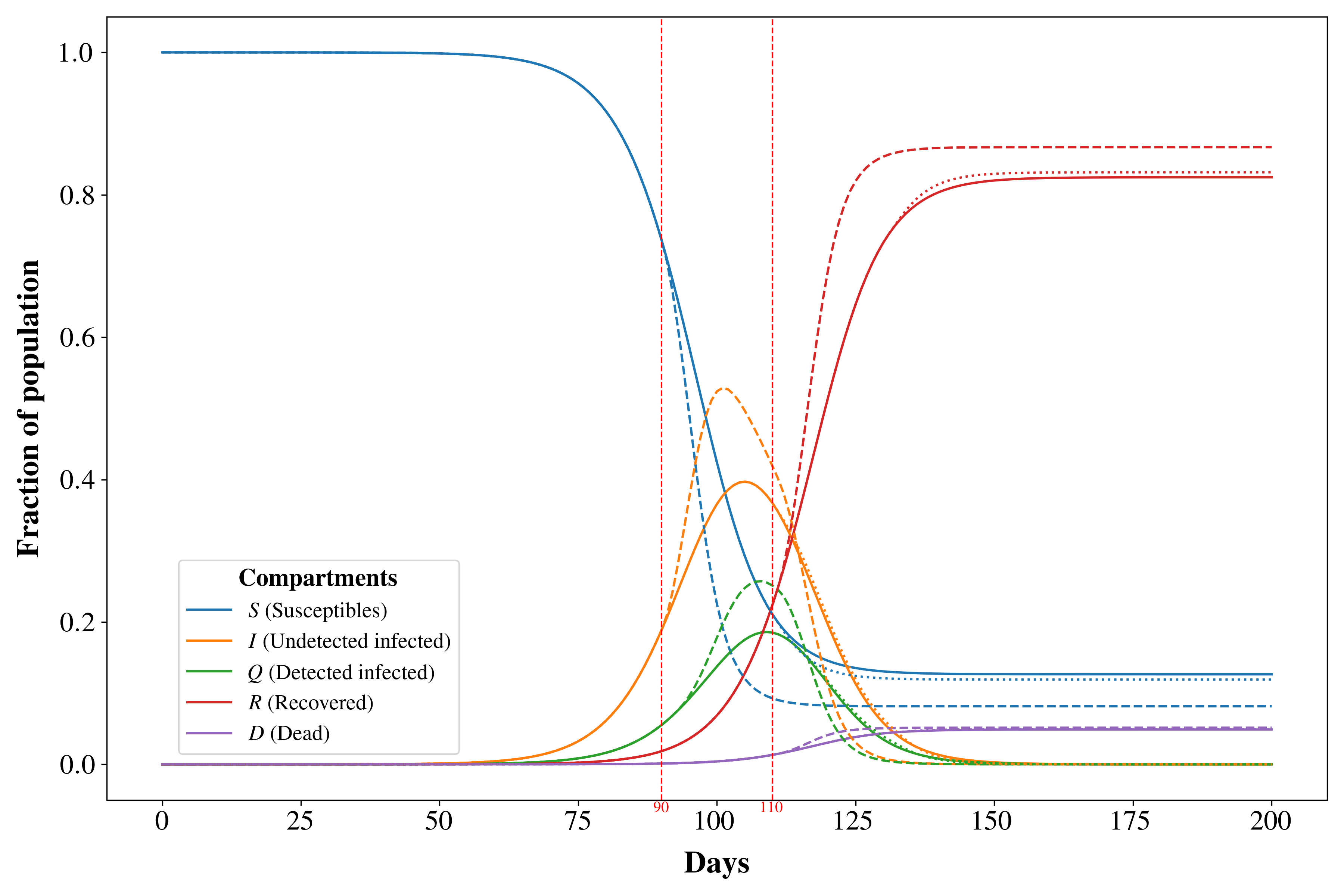}
    \caption{Time evolution of the compartments where the generation-time distribution is switched from \(G_{I1}\) to \(G_{I2}\) at \(T=90\)days and \(T=110\)days. The dashed lines show the trajectories when the switch occurs at \(T=90\), and the dotted lines correspond to the switch at \(T=110\).}
    \label{fig:figure4}
\end{figure}

Our model enables us to predict how changes in the generation-time distribution—arising from either viral mutations or public health interventions—can influence epidemic dynamics. To illustrate this, Figure~\ref{fig:figure4} compares three scenarios. In the first (solid line), the generation-time distribution $G_{I1}$ remains fixed throughout the simulation. In the second (dashed line), the distribution shifts to a new one, $G_{I2}$, after 90 days. In the third scenario (dotted line), the shift takes place on day 110, with $G_{I2}$ exhibiting a shorter average generation time, thereby indicating higher infectiousness.

\begin{figure}[H]
    \centering
    \includegraphics[width=1.0\textwidth]{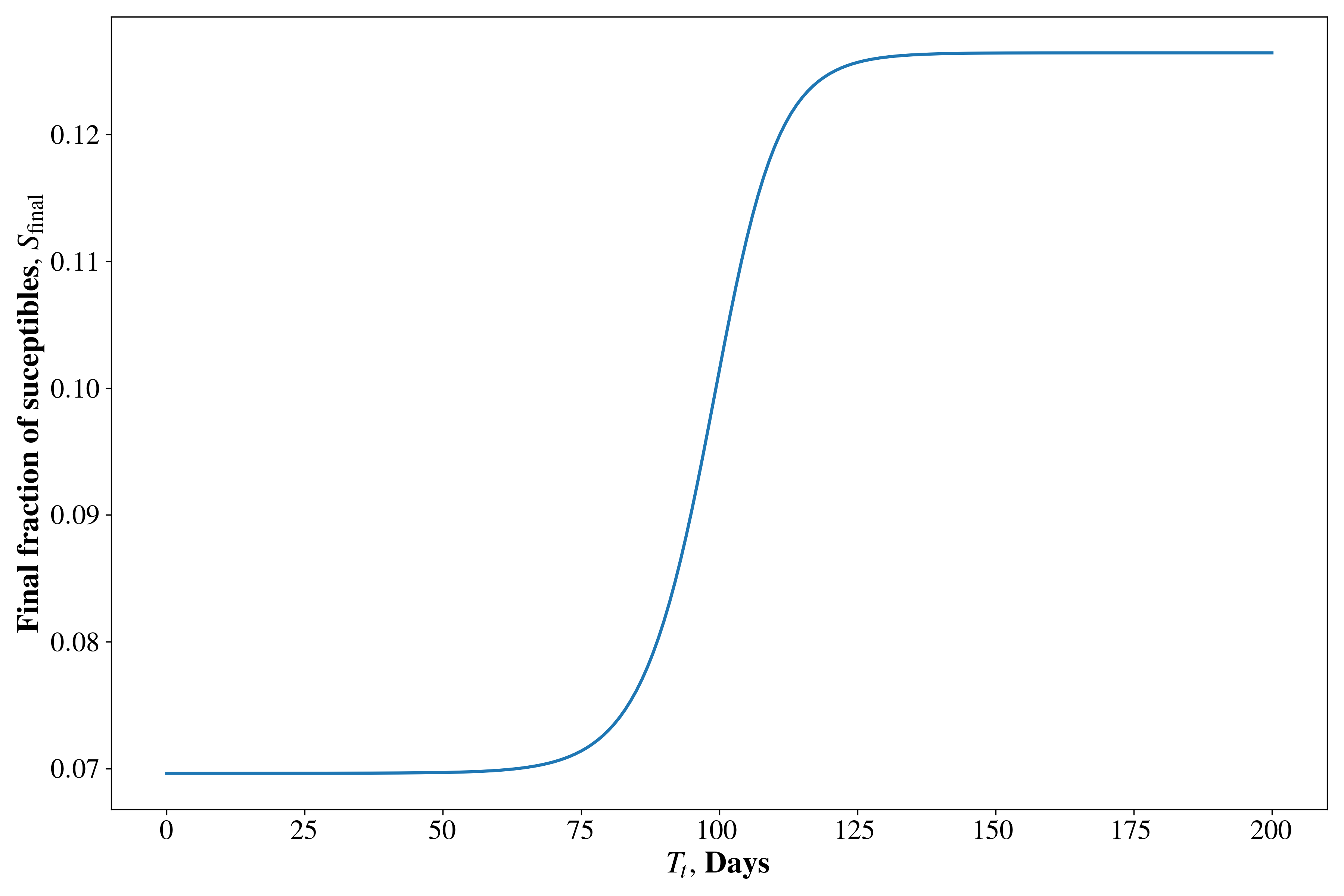}
    \caption{Final fraction of the susceptible population after 200 days of simulation as a function of $T_t$, the time at which the generation-time distribution is switched from $G_{I1}$ to $G_{I2}$.
}
    \label{fig:figure5}
\end{figure}

If the change in the generation time occurs after the infected population has reached its peak, no substantial alteration is observed in the time evolution of the infected population. To illustrate this behavior, we present the plot of $S_{final}$, which represents the final fraction of susceptibles after the end of epidemics (in 200 days, when the stationary regime has already been reached), as a function of the day on which the transition from $G_{I1}$ to $G_{I2}$ takes place (Figure~\ref{fig:figure5}). Although the new generation-time distribution $G_{I2}$ peaks earlier and reflects higher infectiousness, its impact is minimal when the switch happens late: the fraction of unaffected individuals does not change.  Consequently, an increase in infectiousness — or, analogously, a partial relaxation of quarantine measures, such as reducing the duration of isolation — has a little effect on the trajectory of an outbreak if implemented after the peak of infection. This finding is particularly relevant in settings with limited resources, as in many low-income countries, where prolonged quarantines can generate severe collateral consequences, including food shortages among vulnerable populations who depend on daily labor for subsistence.

\begin{figure}[H]
    \centering
    \includegraphics[width=1.0\textwidth]{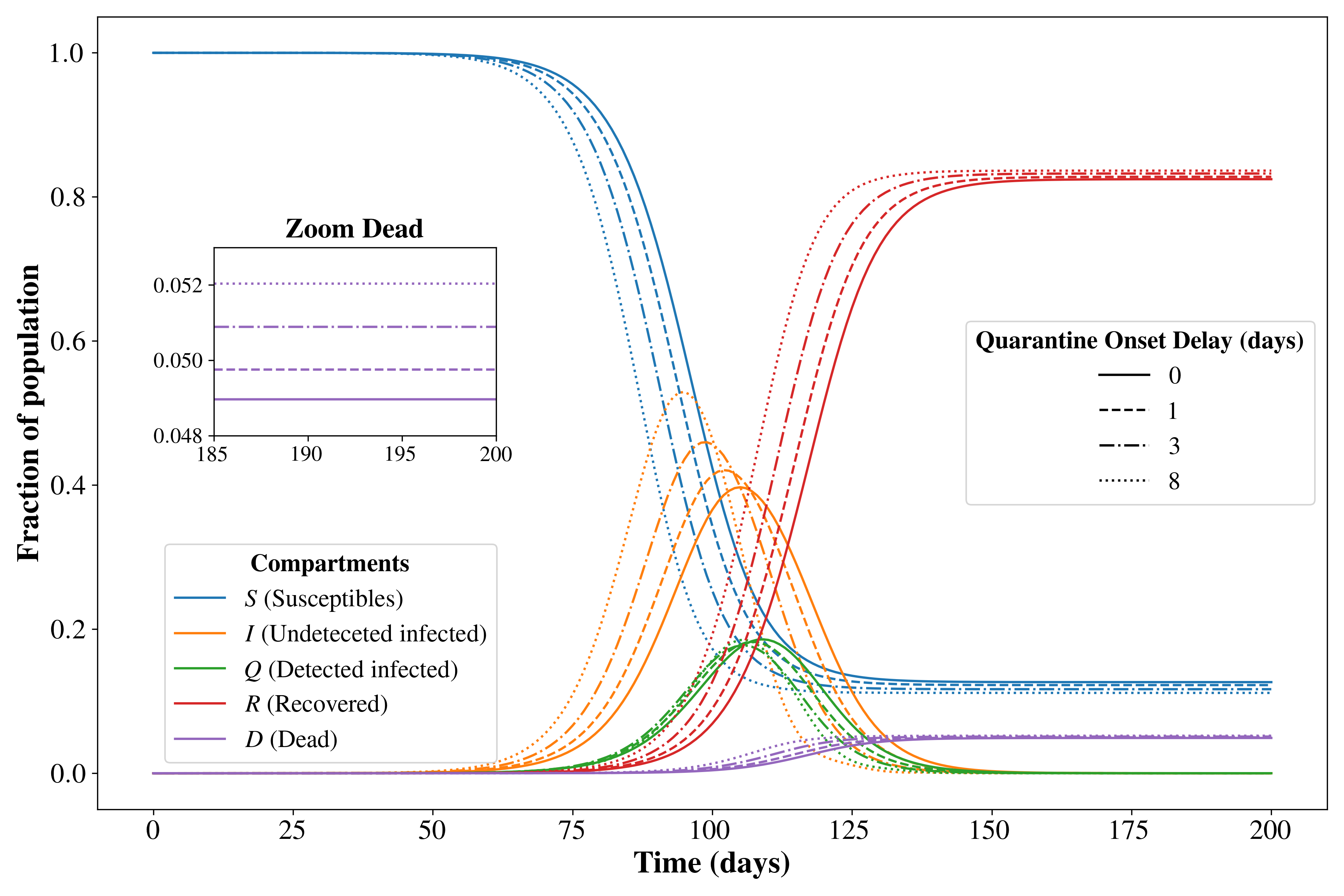}
    \caption{Time evolution of compartments for different delays in the $I \rightarrow Q$ transition.}
    \label{fig:fig6}
\end{figure}

As the COVID-19 pandemic showed, quarantine rules were not always followed to the letter. Many individuals delayed isolating after symptoms began, and even short delays altered how the virus spread. To capture this, our delayed-quarantine model considers what happens if isolation starts 1, 3, or 8 days after symptoms appear. For longer delays, the infection peak appears earlier and sharper (which limits the response options), and the overall mortality increases (Figure~\ref{fig:fig6}).  

The model allows us to quantify how quarantine stringency shapes epidemic dynamics by tracking two key outbreak metrics:
\begin{enumerate}[label=(\alph*)]
    \item \(I_{\text{Max}}\): peak fraction of infected individuals.
    \item \(t_{\text{Max}}\): time at which that peak occurs.
\end{enumerate}

\begin{figure}[H]
    \centering
    \includegraphics[width=0.9\textwidth]{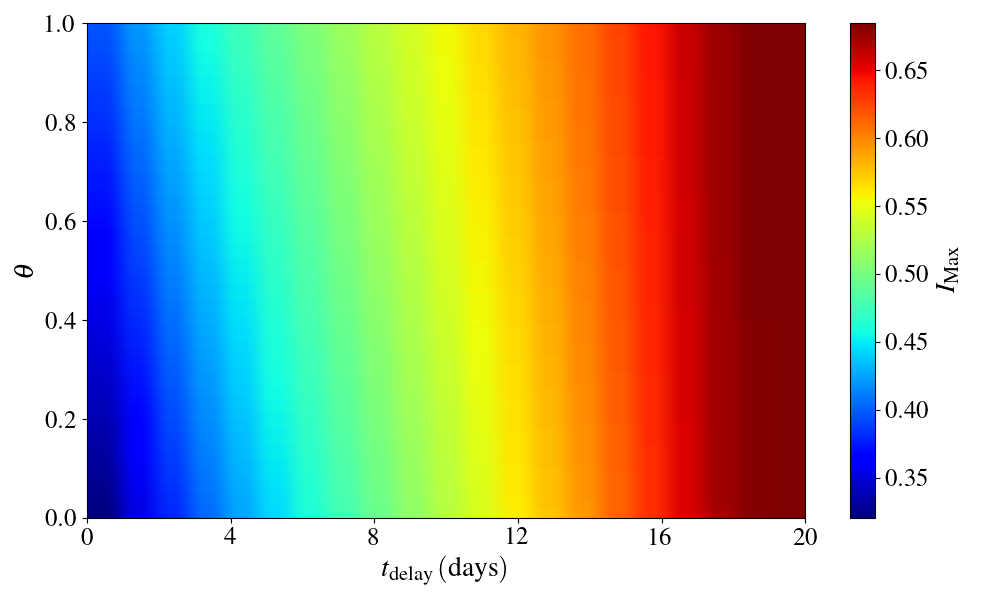}
    \caption{Heatmap of the  the maximum density of infected $I_{max}$ as a function of the quarantine stringency ($\theta$) and the quarantine delay ($t_{\text{delay}}$).}
    \label{fig:fig7}
\end{figure}

\begin{figure}[H]
    \centering
    \includegraphics[width=0.9\textwidth]{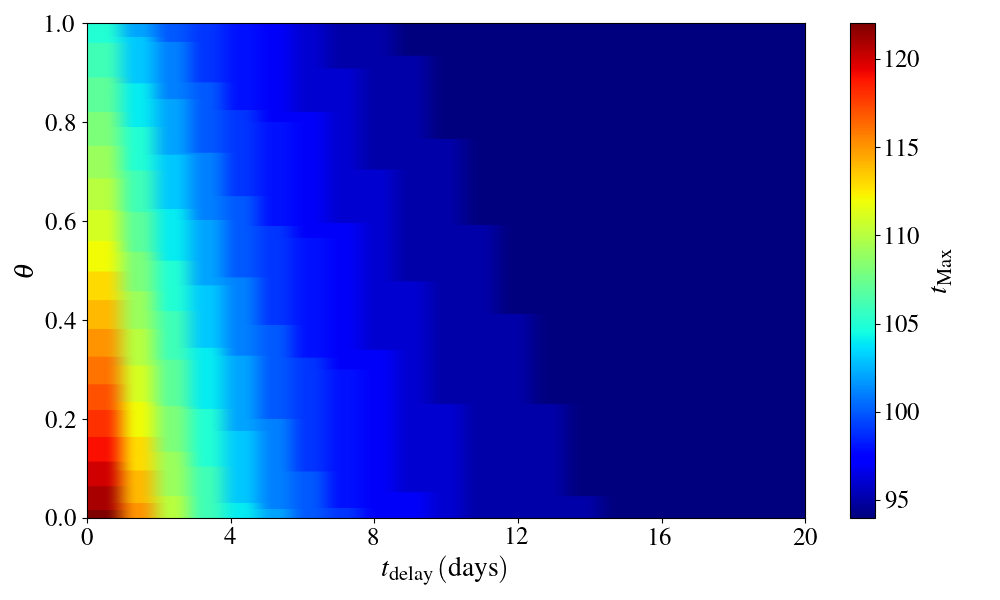}
    \caption{Heatmap of the time when infection peaks $t_{max}$ as a function of  the quarantine stringency ($\theta$ and the quarantine delay ($t_{\text{delay}}$).}
    \label{fig:fig8}
\end{figure}

By jointly varying the stringency parameter \(\theta\) and the quarantine onset delay \(t_{\text{delay}}\), we can compare the effectiveness of alternative intervention strategies. Figures~\ref{fig:fig7} and~\ref{fig:fig8} synthesize these results: Figure~\ref{fig:fig7} presents a heatmap of the peak infection fraction \(I_{\text{Max}}\), which ranges from \(\approx 30\,\%\) under highly stringent, rapid-response conditions to \(\approx 70\,\%\) in the least favorable scenario, while Figure~\ref{fig:fig8} shows that the corresponding peak time \(t_{\text{Max}}\) shifts from about \(120\) days in the most favorable regime to roughly \(95\) days when quarantine is weak and delayed.
 
\begin{figure}[H]
    \centering
    \includegraphics[width=1.0\textwidth]{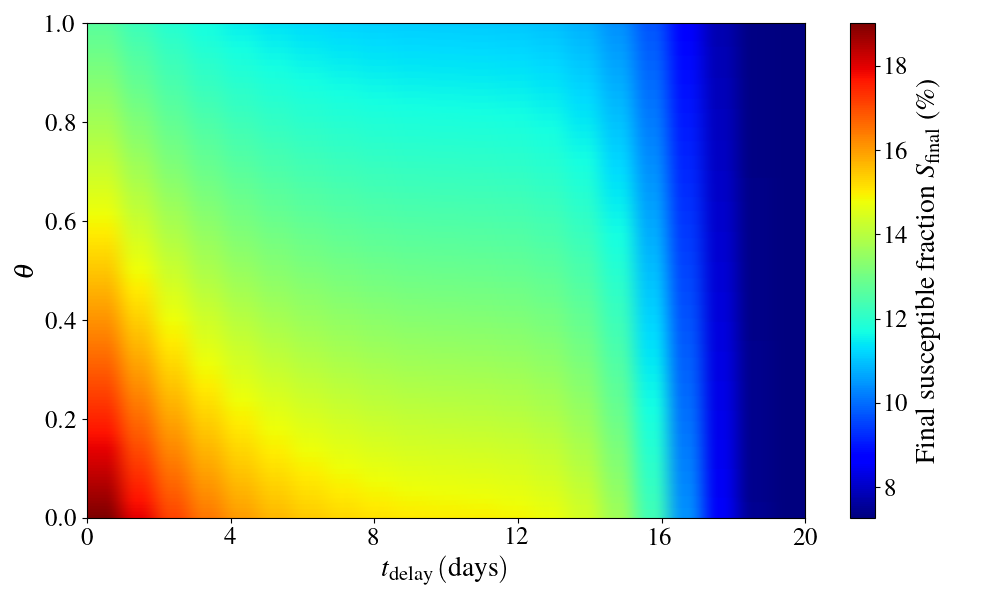}
    \caption{Heatmap of the final fraction of susceptible individuals as a function of the quarantine stringency ($\theta$ and the delay ($t_{\text{delay}}$).}
    \label{fig:fig9}
\end{figure}

A heatmap of the final number of susceptibles (or individuals that remain uninfected by the end of epidemics) as a function of the quarantine stringency~\(\theta\) and delay~\(t_{\text{delay}}\) is shown in Figure~\ref{fig:fig9}. 
Under the most favorable scenario defined by highly stringent isolation measures and rapid entry into quarantine, this final fraction of susceptibles converges to \(\approx 20\,\%\), whereas in the least favorable scenario it decreases to \(\approx 8\,\%\).

\begin{figure}[H]
    \centering
    \includegraphics[width=1.0\textwidth]{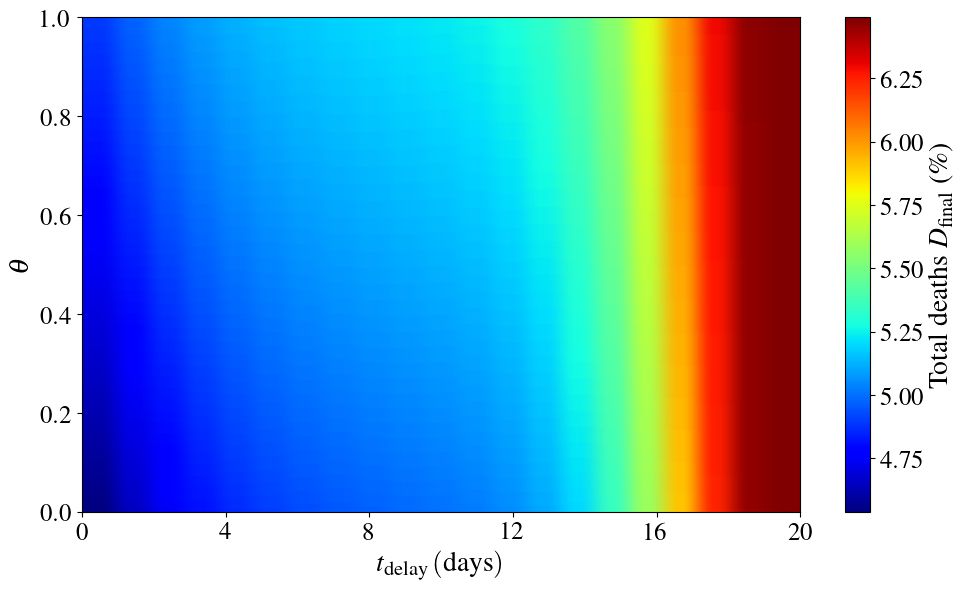}
    \caption{Heatmap of the final fraction of deaths individuals as a function of the quarantine stringency  ($\theta$) and delay ($t_{\text{delay}}$).}
    \label{fig:fig10}
\end{figure}

 Finally, a heatmap of how the fraction of cumulative deaths depends on quarantine stringency~\(\theta\) and delay~\(t_{\text{delay}}\) is presented in Figure~\ref{fig:fig10}. Under the most favorable conditions of highly stringent quarantine measures and rapid entry into quarantine, the cumulative fraction of the population that dies plateaus at \(\approx 4\,\%\); by contrast, in the least favorable scenario, this fraction rises to \(\approx 7\,\%\).
\section{Conclusion}
In this work, we suggest improving current epidemiological models by incorporating the history of infections into population dynamics. The resulting delay equations are only slightly more complex than the traditional SLIR ordinary differential equation models, yet they allow one to make precise and credible predictions of the effects of early detection and isolation of infected individuals, including pre-symptomatic ones, on the spread of infection. This is especially relevant to epidemics like COVID-19, largely propagated by pre-symptomatic individuals. Our approach, fitted to the clinical data and the data from detailed epidemiological case studies, could be incorporated as a more precise single-node dynamic of a well-mixed population into large-scale spatial network-based epidemiological models. 

Numerical experiments with our model yielded two major conclusions: First, we show that relaxing the isolation protocol, when done after the infection peaks, does not noticeably alter the outcome of an outbreak. This observation is particularly relevant for vulnerable populations that depend on daily work for subsistence. Second, it is demonstrated that a stringent quarantine succeeds in mitigating the infection spread only when it is implemented within the first few days after symptom onset. Once this temporal threshold is exceeded, the measure’s effectiveness declines sharply, exhibiting behavior that is practically indistinguishable from the least stringent quarantine scenario. These findings suggest that, in order to curb the rise in new cases, reduce mortality, and preserve a larger share of the susceptible population, the urgency with which individuals enter quarantine after their initial symptoms appear is a decisive factor. Indeed, this aspect turns out to be even more relevant than the degree of quarantine stringency itself.
\printbibliography
\end{document}